\documentclass[
    ,final            % use final for the camera ready runs
  ]
  {aipproc}
\usepackage{amsmath}
\usepackage{smartref}
\layoutstyle{8x11single}
\def\pap{{p\overline{p}}}
\begin{document}

\title{Correlation between mean transverse momentum and\, multiplicity of charged particles in $pp$ and $\pap$  collisions:\newline from ISR to LHC }


\author{E.O.Bodnya}{
  address={University of California, Berkeley, USA}
}

%\author{D.A.Derkach}{
%  address={Oxford University, UK}
%}

\author{V.N.Kovalenko}{
  address={Saint Petersburg State University, Russia}
}
\author{A.M.Puchkov}{
  address={Saint Petersburg State University, Russia}
}
\author{G.A.Feofilov}{
  address={Saint Petersburg State University, Russia}
%,e-mail:{feofilov@hiex.phys.spbu.ru} 
% additional visiting addressl
}

\begin{abstract}
 %%GF
 We present our  analysis  of the available experimental data on correlation between mean transverse momentum and charged particles multiplicity (${\langle p_T \rangle}$-$N_{\text{ch}}$)
%VKov
 at  central rapidity  in $pp$ and $\pap$  collisions at $\sqrt{s}$ from 17 GeV to 7 TeV. A multi-pomeron exchange model based on Regge-Gribov  approach  provides quantitative description of  ${\langle p_T \rangle}$-$N_{\text{ch}}$
%VKov
 correlation data and their energy dependence. Results are found to be in agreement with string fusion model hypothesis.

% A modified  variant \cite{VK-PoS-2013,Vestnik} of multi-pomeron model with  collectivity effects \cite{ADF} is used. The  number of free parameters in the new model is reduced from three to two by fixing a smooth logarithmic growth with energy for the  parameter $k$ -   mean rapidity density of charged particles from one string. Values of parameter $\beta$, that is responsible in an effective way for  quark-gluon string fusion effects,  and of parameter $t$ - string tension,  were derived by fitting the experimental  ${\langle p_T \rangle}$-$N_{\text{ch}}$
%VKov
% correlation data. Smooth behavior of the parameter $\beta$ with energy was obtained that provided possibility of  extrapolation to the highest LHC energy.
% We succeeded also to describe also the experimental data on mean multiplicity. Results on ${\langle p_T \rangle}$-$N_{\text{ch}}$
%VKov
%correlation analysis are found to be in agreement with string fusion model hypothesis.
\end{abstract}

\maketitle

%%%%%%%%%%%%%%%%%%%%%%%%%%%%%%%%%%%%%%%%%%%%
%% MAINMATTER
%%%%%%%%%%%%%%%%%%%%%%%%%%%%%%%%%%%%%%%%%%%%

\section{Introduction}

It came first from the observations in the cosmic rays experiments \cite{cosmic,cosmic-2} and it was confirmed later  in numerous collider studies    that the majority of particles (pions) produced in  high energy $pp$ and $\pap$ collisions   have their transverse momentum $p_T$ as small as a few hundreds of MeV/$c$. The existence of correlation of charged particles mean transverse momentum (${\langle p_T \rangle}$)  with the event multiplicity  ($N_{\text{ch}}$) was established in a wide energy range, changing from the negative (at low collision energies of $\sqrt{s}$=17 GeV) to positive one (above 40 GeV), see a compilation of results in \cite{ADF}.  Besides this, a peculiar dependence of $p_T$ spectra on the event complexity -- a tendency of flattening of the mean transverse momentum of charged particles with the multiplicity --  was established with the growth of  collision energy.  It was argued by Van Hove \cite{VanHove} that this anomalous behavior (a plateau-like structure) of the mean transverse momentum  as a function of hadron multiplicity could be a signal for the deconfinement transition of hadronic matter in $pp$ and $\pap$ high energy collisions.

Recently, the new set of detailed experimental data on the mean transverse momentum ${\langle p_T \rangle}$ versus the charged-particle multiplicity $N_{\text{ch}}$ appeared, showing dramatic difference  of correlation functions  for different colliding systems \cite{alice-ptn}. In $pp$ and $p\text{--}Pb$  collisions at LHC energies, a strong increase of ${\langle p_T \rangle}$ with $N_{\text{ch}}$  is observed, which is much stronger than that measured in $Pb\text{--}Pb$  collisions. In the last case a striking flat behaviour of ${\langle p_T \rangle}$-$N_{\text{ch}}$ correlation was found for the first time  in a wide region of collision centralities. It is discussed in \cite{alice-ptn} that for the $pp$ collisions,  the behavior of ${\langle p_T \rangle}$-$N_{\text{ch}}$ correlation with $N_{\text{ch}}$ could be attributed, within a model of hadronizing strings, to multiple-parton interactions and to a final-state color reconnection mechanism. At the same time it is noted in \cite{alice-ptn} that the data in $p\text{--}Pb$ and $Pb\text{--}Pb$ collisions cannot be described by an incoherent superposition of nucleon-nucleon collisions challenging  most of the event generators.

         However,  quantitative and consistent  description of behavior of ${\langle p_T \rangle}$-$N_{\text{ch}}$ correlation in nucleon-nucleon  collisions and, in particular,  its energy dependence, still remains an open basic problem. It is evident that the last one should be completely understood, on a single approach  base, before any use of event-generator calculations for the more complicated  cases of $p\text{--}Pb$ or $Pb\text{--}Pb$ collisions analysis. Overview of different models trying to explain the effects observed in  ${\langle p_T \rangle}$-$N_{\text{ch}}$ correlations in  high energy $pp$ and $\pap$ collisions could be found in \cite{ADF}.

         In the present study we continue the analysis \cite{VK-PoS-2013,Vestnik} of ${\langle p_T \rangle}$-$N_{\text{ch}}$ correlation  for charged particles using available data in a wide energy range of  $pp$ and $\pap$  collisions and basing on the Regge-Gribov approach. It is assumed \cite{Capella1}, \cite{KTM82} that low-$p_T$ multiparticle production
% in peripheral collision of hadrons
is a result  a few ($n$) pomeron exchange. After this exchange hadrons become joined by $n$ pairs of strings. Fission of the strings during the separation of hadrons results in production of $n$-pomeron showers, so that the  multiplicity exceeds by a factor $n$ the one-pomeron shower multiplicity. In case of identical strings the $p_T$ distributions of charged particles have no dependence on  multiplicity.  However, when there is possible interaction between strings \cite{abramovskii} one may expect the $p_T$ distributions of charged particles to  be different from those  produced  by non-interacting strings. It is in the case of interaction between quark-gluon strings in a form of color strings fusion \cite{BP} that new types of particle-emitting sources (strings) might be formed, characterized by a higher energy density (described by string tension parameter $t$). As a result, the higher values of mean transverse momenta of charged particles could be expected in hadronization of such types of strings.

         We are using a  modified  variant \cite{VK-PoS-2013,Vestnik} of multi-pomeron model \cite{ADF} with  collectivity effects. Similar to \cite{ADF}, both multiplicity of charged particles and their transverse momenta are treated within the same model where the multi-pomeron exchange is combined with Schwinger mechanism \cite{schwinger} of particle production. Thus  the collectivity effects like string fusion are included into the model \cite{ADF} in an effective way by a single parameter $\beta$, and the ${\langle p_T \rangle}$-$N_{\text{ch}}$ correlation appears to be the intrinsic property of the model.

%%GF
 %It is shown below that this concept realized in the  modified, 2-parameter  variant of multi-pomeron exchange  model \cite{VK-PoS-2013,Vestnik} allows to provide  good quantitative description of ${\langle p_T \rangle}$-$N_{\text{ch}}$ correlation pattern and various observables in $pp$ and $\pap$ collisions and their energy dependence in a wide energy range.
  It is shown below that this concept
  %realized in the  modified, 2-parameter  variant of multi-pomeron exchange  model \cite{VK-PoS-2013,Vestnik}
  allows to provide  good quantitative description of ${\langle p_T \rangle}$-$N_{\text{ch}}$ correlation pattern and various observables in $pp$ and $\pap$ collisions and their energy dependence in a wide energy range.
  %%GF
  %Available data on ${\langle p_T \rangle}$ and ${\langle{N_{\text{ch}}}\rangle}$ vs. collision energy as well as multiplicity distributions
          %and ....
 %are  well-described assuming a  mechanism of string fusion as the one responsible for the ${\langle p_T \rangle}$  plateau with $N_{\text{ch}}$.
 It is also shown that the model has the predictive power that allows extrapolation to the higher collision energies.
 %%GF
  Besides, the model could be extended to the long-range ${\langle p_T \rangle}$ and ${\langle{N_{\text{ch}}}\rangle}$ correlations which might be studied by using event-by-event data in separated pseudorapidity intervals.

         The paper is organized as follows: in the next section we give a very brief description of the  modified  variant of multi-pomeron exchange model
         %%GF
         with collectivity
         used in our analysis. In Section 3 we describe how the values of the parameters in the model are extracted by fits to experimental data. In Section 4 we discuss the results obtained and show some predictions. Finally, we present our conclusions.

\section{Modified multi-pomeron exchange model % \cite{VK-PoS-2013,Vestnik}
}

In the present study we use the modifications of the multi-pomeron exchange model  with string fusion \cite{ADF} as described in \cite{VK-PoS-2013,Vestnik}. We consider the multiplicity and the transverse momenta of charged particles produced in high energy  nucleon-nucleon collisions into a given pseudo(rapidity) interval $\delta.$

%%GF
Basing on \cite{KTM82}, \cite{Kaidalov-87} and following \cite{ADF}, the function $f(N_{\text{ch}}, p_T; z; k, \beta, t)$ is introduced to describe both multiplicity and transverse momentum distributions in soft multi-particle production in  hadron collisions:

\begin{equation}\label{f-distribution}
f(N_{\text{ch}}, p_T; z; k, \beta, t)={C_w}
\sum\limits_{n=1}^{\infty} \dfrac{1}{z n}\left( 1- \exp(-z)\sum\limits_{l=0}^{n-1}\frac{z^l}{l!}\right) \cdot
\text{exp}(-2nk\delta)\dfrac{(2nk\delta)^{N_{\text{ch}}}}{N_{\text{ch}!}}\cdot
\dfrac{1}{n^\beta t} \text{exp} \left(-\dfrac{\pi p_T^2}{n^\beta t} \right).
\end{equation}
Here $C_w$ is a normalization factor which is introduced in order to
enforce the normalization  condition:

  \begin{equation}
    \label{v2}
    2 \pi
    \sum
    \limits_{N_{\text{ch}}=0}^{\infty} \
    \int
    \limits_{0}^{\infty}
    {
    f(N_{\text{ch}}, p_{T}; z; k, \beta, t)
    p_{T}dp_{T}
    }
    =
    1\,.
  \end{equation}

The function $f$ (see eq.(\ref{f-distribution})) represents combined probability distribution of the number of charged particles $N_{\text{ch}}$ and their transverse
momentum spectra ($p_T$). The first factor in eq.(\ref{f-distribution}) gives a probability of $n$ pomerons production in a single event \cite{KTM82,Kaidalov-87}. Parameter $z$ is responsible for the
dependence of this distribution on collision energy $\sqrt{s}$.
%%GF We used the following values of the parameters
We use the following values of the parameters \cite{smth}:
\begin{equation}
z=\frac{2C\gamma s^\Delta}{R^2+\alpha'\log(s)},\
\Delta = 0.139,\  \alpha'=0.21 \text{ GeV}^{-2},\
 \gamma=1.77 \text{ GeV}^{-2},\
  R_0^2=3.18 \text{ GeV}^{-2},\  C=1.5\ .
\end{equation}

The second multiplier in eq.(\ref{f-distribution}) represents the probability of $N_{\text{ch}}$ particles production in $n$-pomeron showers. The Poisson distribution is used with mean being proportional to the number of pomerons $n$ and the width of pseudo(rapidity) interval $\delta$. Parameter $k$ is, accordingly, the mean
number of charged particles per rapidity unit from one string.

%%GF
The last factor  was introduced in eq. (\ref{f-distribution}) in \cite{ADF}. It reflects Schwinger mechanism of particle production \cite{schwinger}, which prescripts
the transverse spectrum of charged particles from one string to be of a  Gaussian form:
\begin{equation}\label{Schwinger}
\frac{d^2 N_{\text{ch}}} {dp_{T}^2} \sim
\text{exp} \left( \frac{-\pi(p^2_t+m^2)}{t} \right),
\end{equation}
where $t$ corresponds to the string tension.

%%GF Note that in our approach a simplified assumption is used: we do not discriminate the particle species, considering only production of pions, that eliminates the particle mass dependence in the expression (\ref{f-distribution}).
%%GF
 A simplified assumption is used in our work: we do not discriminate the particle species, considering only production of pions.   Thus the particle mass dependence in the expression eq.(\ref{f-distribution})
is  eliminated. %VKov

In the current model, similar to \cite{ADF}, the string interaction is introduced in an effective way through a single parameter $\beta$ responsible for the collectivity effects, that gives us an effective string tension $n^\beta t$.

$\langle p_T \rangle\text{-} N_{\text{ch}}$ correlation function in the present model is calculated as

\begin{equation}\label{pt-Nch-function}
   \langle p_{T} \rangle_{N_{\text{ch}}}=\dfrac{\int\limits_0^\infty
   {f(N_{\text{ch}}, p_T; z; k, \beta, t) p_T^2 dp_{T}}} {\int\limits_0^\infty
   {f(N_{\text{ch}}, p_T; z; k, \beta, t) p_T dp_T}},
\end{equation}
or
\begin{equation}\label{pt-Nch-function-dec}
   \langle p_{T} \rangle_{N_{\text{ch}}}=\frac{1} {\int\limits_0^\infty
   {f(N_{\text{ch}}, p_T; z; k, \beta, t) p_T dp_T}} \sum\limits_{n=1}^{\infty} \int\limits_0^\infty
   {f_n(N_{\text{ch}}, p_T; z; k, \beta, t) p_T^2 dp_{T}},
\end{equation}
where $f_n(N_{\text{ch}}, p_T; z; k, \beta, t) p_T^2 dp_{T}$ here is defined similar to eq.(\ref{f-distribution}), but without the sum symbol.
%\frac{1} {\int\limits_0^\infty
%   {f p_T dp_T}}
%   {
%\sum\limits_{n=1}^{\infty} \frac{{C_w} }{z n}\left( 1- e^{-z}\sum\limits_{l=0}^{n-1}\frac{z^l}{l!}\right) \cdot
%e^{-2nk\delta}\dfrac{(2nk\delta)^{N_{\text{ch}}}}{N_{\text{ch}!}}\cdot
%\dfrac{1}{n^\beta t}
%\int\limits_0^\infty{ e^ {-\dfrac{\pi p_T^2}{n^\beta t}} p_T^2 dp_{T}}}
%.

Note, that it is possible to integrate out $p_T$ dependence from eq.(\ref{f-distribution}) and obtain the charged particles distribution and mean multiplicity:

%%GF
\begin{equation}\label{e2}
  {P}(N_{\text{ch}}) =
  2 \pi
  \int
  \limits_{0}^{\infty}
  {f(N_{\text{ch}}, p_T; z; k, \beta, t) p_{T} dp_{T}}, \hspace{0.6cm}
   %\langle %VKov
   %N_{\text{ch}}
   %\rangle (s) =
   %\sum
   %\limits_{N_{\text{ch}}=0}^{\infty}
   %{
   %N_{\text{ch}}
   %{P}(N_{\text{ch}})
   %}
  %= 2 n k \delta.
\end{equation}
%%
%%GF
\begin{equation}\label{e2-a}
   \langle
   N_{\text{ch}}
   \rangle (s) =
   \sum
   \limits_{N_{\text{ch}}=0}^{\infty}
   {
   N_{\text{ch}}
   {P}(N_{\text{ch}})
   }
  = 2  \langle
   n(s)
   \rangle k(s) \delta. %VKov
\end{equation}

The dependence of the mean transverse momentum of charged particles on energy can be calculated using:

  \begin{equation}
   \label{a9}
   \langle
   p_{T}
   \rangle (\sqrt{s}) =
   \dfrac
   {
   \sum
   \limits_{N_{\text{ch}}=0}^{\infty}
   {
   N_{\text{ch}}
   \int
   \limits_{0}^{\infty}
   {f(N_{\text{ch}}, p_T; z; k, \beta, t)
   p_{T}^{2} dp_{T}}}
   }
   {
   \sum
   \limits_{N_{\text{ch}}=0}^{\infty}
   {
   N_{\text{ch}}
   \int
   \limits_{0}^{\infty}
   {f(N_{\text{ch}}, p_T; z; k, \beta, t)
   p_{T} dp_{T}}}
   } =
   2 \pi
   \sum
      \limits_{N_{\text{ch}}=0}^{\infty}
      {
   \dfrac
{
   N_{\text{ch}}
}
   {
   \langle
   N_{\text{ch}}
   \rangle %(s)
   }
   \int
   \limits_{0}^{\infty}
   {f(N_{\text{ch}}, p_T; z; k, \beta, t)
   p_{T}^{2} dp_{T}}
   }
   \,.
  \end{equation}

One should mention, that the picture where transverse momentum
of charged particles increases with multiplicity and collision energy
is natural for
%%GF so-called
a string fusion model \cite{StringFusion0, StringFusion}.
According to this model, with the increase of the collision energy one may expect a general growth in the  number of multi-pomerons in exchange in nucleon-nucleon collision. Thus the total number of strings will increase and they could start to overlap, forming clusters.  In case of  string fusion  one may expect the formation of new sources with higher string tension and, therefore,   the increase of both mean multiplicity and mean $p_T$ of charged particles emitted from such type fused strings. It could cause non-zero correlation
between transverse momentum and multiplicity with allowance to both
positive \cite{smth,VLP,VKYaF} and negative \cite{NA61Negative,AndronoPoS}
$\langle p_T \rangle\text{-} N_{\text{ch}}$
 correlations.

%%GF
This increase in mean $p_T$  is relevant to the degree of strings overlap in transverse space that is characterized by the string density parameter  \cite{StringFusion0,StringFusion} which is denoted here as $\eta_t$. According to string fusion model, the mean multiplicity $\mu$ and mean value of $p_T^2$ emitted from a cluster of fused strings are modified compared to multiplicity $\mu_0$ and transverse momentum squared $p_0^2$ of a single string:

  $$
   \mu=\mu_0 \sqrt{\eta_t}, \hspace*{1cm} p_T^2=p_0^2  \sqrt{\eta_t}.
  $$
 Here $\mu_0$ and $p_0^2$ are treated as some universal, independent on energy parameters, as they correspond to the properties of single (none-fused) string.
  Thus one can obtain the following ratio, which in case of the string fusion we should expect energy independent:

    \begin{equation}
   \label{b5}
   \frac{\mu}{p_T^2}=
   \frac{\mu_0}{p_0^2}
  \end{equation}

 In this connection, we have to note
 %%GF here
 that it is possible to obtain this ratio with two basic quantities of our model: $k$ (mean number of particles produced per unit rapidity by one emitter) and the mean value of squared transverse momentum
%, which is  approximately
  $\sim{\langle n \rangle}^\beta t$.
So the both quantities, as in eq.(\ref{b5}), could be defined using the ${\langle p_T \rangle}$-$N_{\text{ch}}$ correlation function at the given $\sqrt{s}$.
 Therefore, the condition eq.(\ref{b5}) %obtained ,
 could be checked
in the framework of the present model
  by plotting the values
%$R=k/({\langle n\rangle}^\beta t)$
\begin{equation}
   \label{R1}
   R=\frac{k}{{\langle n\rangle}^{\beta} {t}}
\end{equation}
as a function of collision energy using 2 parameters extracted from the experimental data at the given $\sqrt{s}$.

\section{Determination of model parameters}
\subsection{Determination of the energy dependence of the parameter $k$ }

\begin{figure}
  \includegraphics[height=.3\textheight]{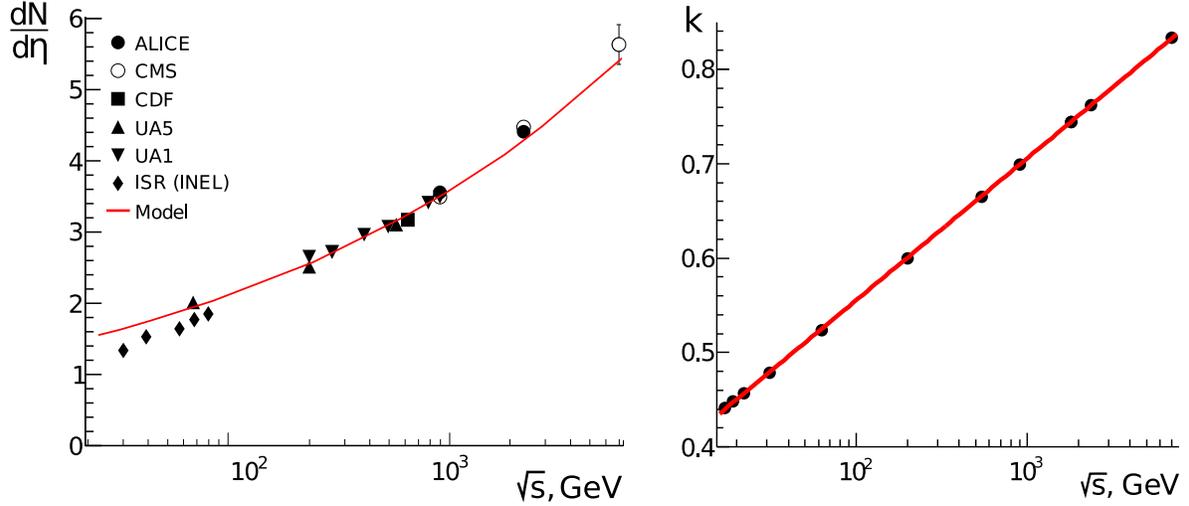}
  \label{mult}
  \caption{Left: Experimental data on charged particle pseudorapidity density at midrapidity  vs. collision energy (see compilation of pp and $\pap$ collision data in \cite{abelev}  and the modified multi-pomeron exchange model -- solid line
  %%GF
  (see text).
  Right: Energy dependence of parameter $k$ obtained in the present model using data at different energies. Straight line -- approximation by $k = 0.255+0.0653 \log{\sqrt {s}}$.}
\end{figure}

 In our modified  variant \cite{VK-PoS-2013,Vestnik}, contrary to \cite{ADF}, $k$ -- the mean number of particles produced per unit of rapidity  per string --  is not considered as a free model  parameter but assumed to be the energy dependent: $k$=$k(\sqrt{s})$. This energy dependence was fixed in the model by fitting a set of available experimental data on charged particles yields vs. $\sqrt{s}$. We used the experimental data on  the energy dependence of the charged particle multiplicity density in $pp$ and $\pap$ collisions (see compilation in \cite{abelev}). The  parameter $k$ was determined by fitting the data in the energy range from {20 GeV} to {7 TeV}, the result of fitting is shown in the Fig. \ref{mult} (left) by solid line.
 %%GF The values of the  parameter $k$ are plotted as a function of $\sqrt{s}$ in Fig.\ref{mult} (right).

 %%GF
 Charged particles pseudorapidity density in the left  plot Fig.\ref{mult} was calculated basing on the eq. \ref{e2-a},
%VKov
 where the mean number of pomerons $\langle n(s)\rangle$ was estimated at a given $\sqrt{s}$  using Regge-Gribov approach, and the pseudorapidity width $\delta$ was selected to match the relevant experimental one.
 Thus the parameter $k$ was defined by fitting experimental data (see the left  plot Fig.\ref{mult}).
 %%GF
 We have to note here that the ISR data (diamonds  in the left  plot Fig.\ref{mult}) were not used in our fitting procedure, so that the results of extrapolation of our model  are shown in this region.

%\begin{figure}
  %\includegraphics[height=.3\textheight]{golfer}
%  \includegraphics[height=.3\textheight]{fig_8}
%  \label{k-fig}
%  \caption{Picture 8 Energy dependence of parameter $k$ obtained in the present model. Dots - results of fits using %data of the Fig.\ref{mult}. Straight line - approximation by $k = 0.255+0.0653 ln\sqrt)s)$. }
%\end{figure}

As a result of this fitting parameter $k$ was defined.
%%GF
The values of the  parameter $k$ are plotted as a function of $\sqrt{s}$ in Fig.\ref{mult} (right).
 A smooth logarithmic growth of multiplicity density from one string with energy was obtained as:  $k = 0.255+0.0653 \log\sqrt{s}$. This result is in agreement with string fusion model predictions \cite{StringFusion0,StringFusion}, where the growth of string tension with the collision energy is expected for fused strings   due to the growing string density.

\subsection{Determination of the energy dependence of the parameters $t$  and $\beta$}

\begin{figure}
  \includegraphics[height=.3\textheight]{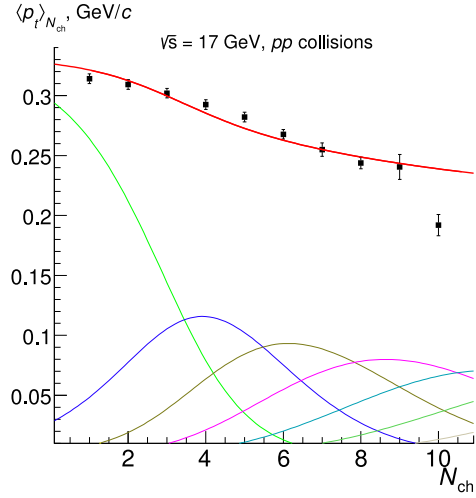}
   \label{ptn17}
   \caption{${\langle p_T \rangle}$-$N_{\text{ch}}$ correlation function at $\sqrt{s}$=17 GeV in a window of 1.5 rapidity units.  Data  from \cite{ptn17}. Curve -- results of fit in modified multi-pomeron exchange model, using eq. (\ref{pt-Nch-function}), to define parameters  $\beta$ and $t$. Contributions of several first terms ($n$=1, 2, 3...) of processes going via the exchange of $n$  soft pomerons are also shown.}
\end{figure}

\begin{figure}
  \includegraphics[height=.3\textheight]{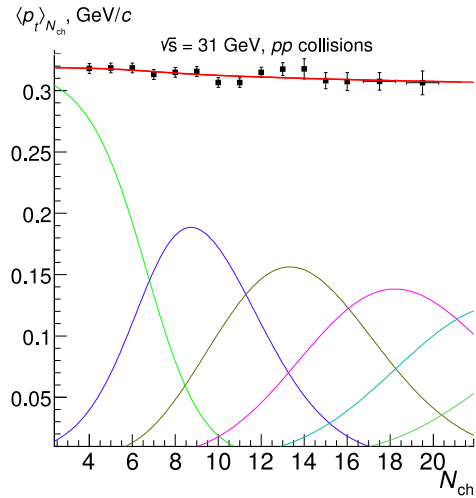}
   \label{ptn31}
   \caption{...the same as in Fig.\ref{ptn17} for  $\sqrt{s}$=31 GeV.  Data - from \cite{ptn31}, $|y|<2$.}
\end{figure}

\begin{figure}
  \includegraphics[height=.3\textheight]{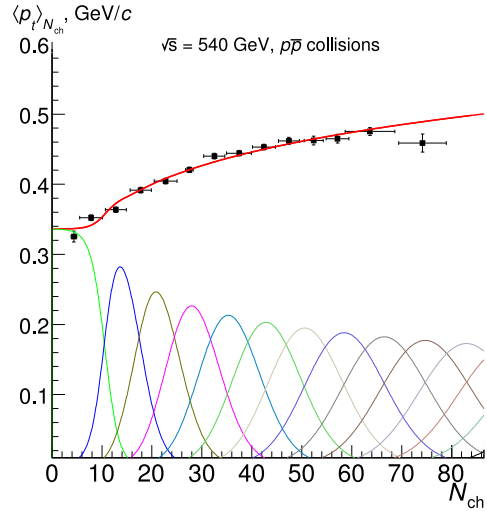}
   \label{ptn540}
   \caption{...the same as in Fig.\ref{ptn17} for  $\sqrt{s}$=540 GeV.  Data  from \cite{ptn540}, $|\eta|<2.4$ }
\end{figure}

\begin{figure}
  \includegraphics[height=.3\textheight]{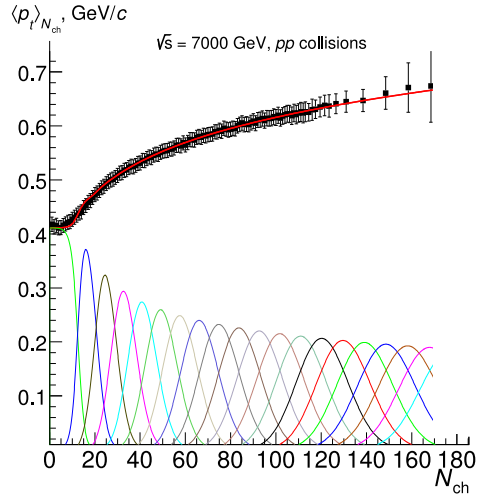}
   \label{ptn7}
   \caption{...the same as in Fig.\ref{ptn17} for  $\sqrt{s}$=7 TeV.   Data  from CMS \cite{DistCMS}. Midrapidity, pseudorapidity  interval is $|\eta|<2.4$.}
\end{figure}

  Two free parameters of the modified model,  $\beta$ - that takes account of string fusion,  and the average string tension parameter $t$, are extracted in our work from the available
  %%GF
  experimental
  data on  ${\langle p_T \rangle}$-$N_{\text{ch}}$ correlation in proton-(anti)proton collisions at wide energy region from {17 GeV} to {7 TeV}.  The expression
  for the correlation function eq.\ref{pt-Nch-function} was used in our fitting procedure.
   Some results of fitting of ${\langle p_T \rangle}$-$N_{\text{ch}}$ correlation in $\pap$ and $pp$ collisions at different energies, starting from 17 GeV  and to 7 TeV, are shown in Fig.  \ref{ptn17} -- \ref{ptn7} (a full data set is available in \cite{VK-PoS-2013}).

%Charged-particles multiplicity distribution obtained in experiment at .... GeV  - 7 TeV  and  the results of the the  modified multi-pomeron exchange model for processes going via the exchange of n  soft pomerons in pp collisions at $\sqrt{s}$ = ... TeV. Data - see ...

%  Mean pT  vs. $\sqrt{s}$

\begin{figure}
  \includegraphics[height=.29\textheight]{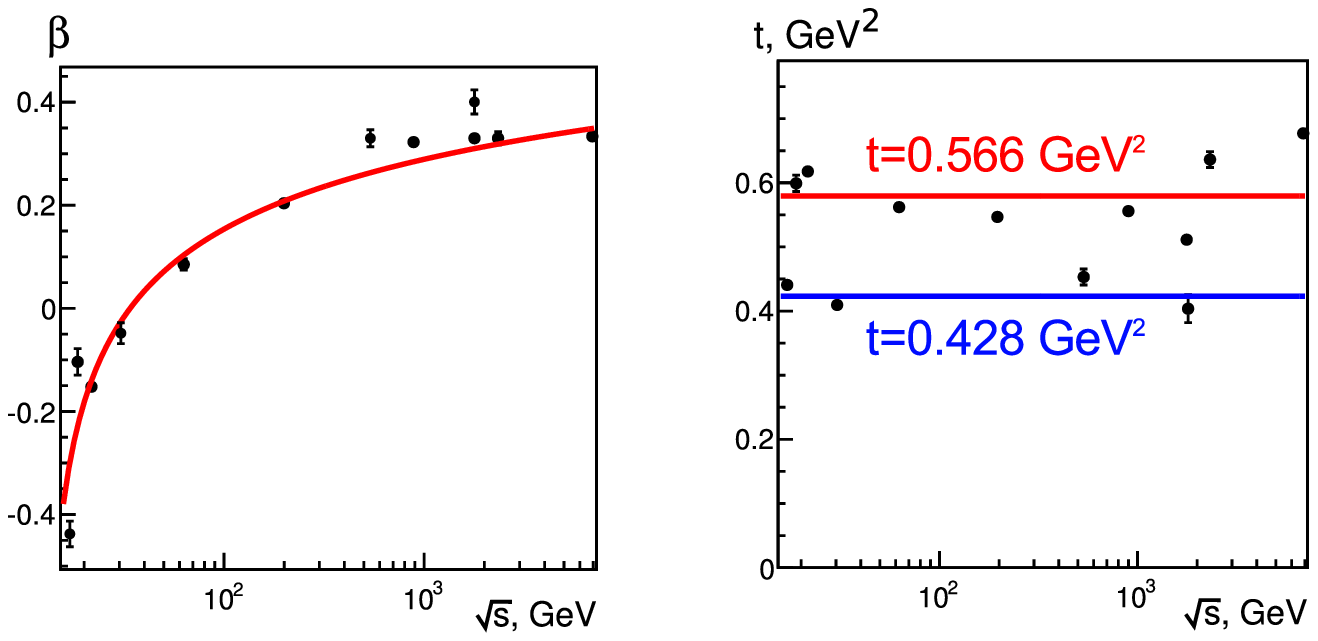}
   \label{beta-t}
   \caption{Energy dependence of parameters $\beta$ and $t$ extracted by fitting experimental data on ${\langle p_T \rangle}$-$N_{\text{ch}}$ correlation.} %VKov
\end{figure}

   %%GF  Extracted values of parameters $\beta$ and $t$ are shown as function on energy at the Fig. \ref{beta-t}.
   Parameters $\beta$ and $t$ were extracted by fitting  ${\langle p_T \rangle}$-$N_{\text{ch}}$ correlation data, the relevant values   are shown as function on energy at the Fig. \ref{beta-t}.
   Both parameters are showing smooth behavior with the collision energy.
   % (see Fig.\ref{beta-t}).
  %%GF Similar to \cite{ADF} we observe here two families of parameter $t$ that points to the systematic effects in
  %the experimental data ( explained in \cite{ADF}  as a result of the different estimate of the mean $p_T$ values for a single pomeron exchange - i.e. for low-multiplicity events). The upper value of string tension is used here: t=
   %(t= in \cite{ADF})

%%GF

   For the parameter $t$
two subsets are obtained: $t$ = 0.566 $\text{GeV}^2$ and $t$ = 0.428 $\text{GeV}^2$ which are very close to those obtained in \cite{ADF}.  As it was explained previously \cite{ADF}, this  fact points to the systematic effects of the different experimental estimates of the mean $p_T$ values for low-multiplicity events (i.e.for  a single pomeron exchange region). The upper value of string tension is used here $t$ = 0.566 $\text{GeV}^2$ as the one   ensuring the values of event mean $p_T$ values corresponding to the experiment (see Fig. \ref{mean-pt-energy} below). %VKov

%t = 0.568 +- 0.001 $\text{GeV}^2$,
%
%t = 0.406 +-  0.006 $\text{GeV}^2$,
%
%in Russian version we have:
%
%t = $(0. 566 \pm 0. 003)$ $\text{GeV}^2$ ,
%
%t= $(0. 428 \pm, 0.005)$ $\text{GeV}^2$ .

 %%GF This discrepancy between two values of t may be caused
%by different procedures of data processing and interpolation to the most soft part, performed by
%various experiments. Most of the point belong to the first subset, and only they are used for further
% analysis.

   %%GF The first subset is used due to the complete correspondence to the mean pT data (similar to \cite{ADF})

   In case of $\beta$
   A smooth behavior of parameter $\beta$ with energy is obtained and approximated at fig. \ref{beta-t}
   by $\beta = \beta_0 ((1-{\log \sqrt{s}-\beta2)}^\beta 1)$ .
  Here we have $\beta_0$ = 1. 16 ± 0. 39 ,
$\beta_1$ = $0. 19 \pm0. 08 $,
$\beta_2$ = $2.52 \pm 0. 03 $.

%%GF
%   As a result of fitting we obtained previously  the energy behavior of $\beta$ in \cite{ADF} ???
% $\beta$  = ${-1.0 \pm 0.2}({\log{\sqrt{s} - (0.14 \pm 0.1)}^{1.9\pm0.2}})  + (0.49\pm0.02)$,

\section{Results and discussion}

\subsubsection{Multi-pomeron exchange  contributions vs. $\sqrt{s}$}

\begin{figure}
  \includegraphics[height=.27\textheight]{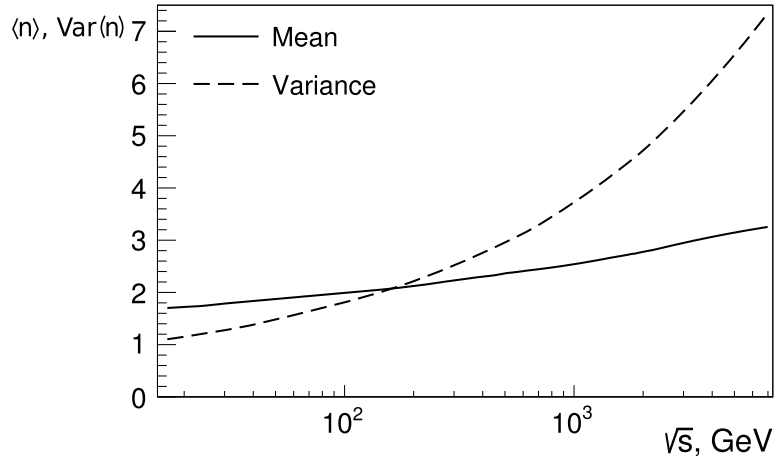}\,\,\,\,\,\,\,\,\,\,\,\,\,\,\,\,\,\,\,\,\,\,\,\,\,\,\,\,\,\,\,\,\,\,\,%VKov %For centering
  \label{pomerons}
  \caption{Results of calculations  of  the  mean number of pomerons (solid line) and of  variance
%VKov
   of  the  mean number of pomerons   vs. $pp$ collision energy.  }
\end{figure}

Fig. \ref{pomerons} contains results of our calculations   of  mean number of pomerons (solid line) and their variance  vs. collision energy.
%Calculations are done with two model parameters $\beta$ and $t$ as defined in the previous section by fitting the available experimental data on ${\langle p_T \rangle}$-$N_{\text{ch}}$ correlation.
 One may see a rather  smooth growth of  the  mean number of pomerons with energy, reaching the value about 3 at the LHC region. This increase is combined with a much faster growth of variance of  the  mean number of pomerons.

Contributions to the correlation function of several first  terms ($n$ = 1, 2, 3  etc. -- see formula (\ref{pt-Nch-function-dec})) are shown at various energies in the Figures   \ref{ptn17} - \ref{ptn7}. One may see that the dominant contribution of single pomeron exchange is concentrated in the low multiplicity region
%(below Nch ~10)
, bringing there the values of ${\langle p_T \rangle}$  from $\sim$ 0.32 GeV/$c$ to 0.4 GeV/$c$ (for the relevant collision energies  of 17-7000 GeV, see Figures \ref{ptn17} and \ref{ptn7}).

With the growing energy the role of the multi-pomeron exchange increases. The number of pomerons  and fluctuations in the number of pomerons are growing  with energy leading to the observed flattening of the ${\langle p_T \rangle}$ - $N_{\text ch}$ correlation functions.

We have to note here that the growth of fluctuations in the number of pomerons exchanged in the collision of protons at the LHC energies brings relevant large fluctuations in the number of particle-emitting strings and, therefore, has a natural prediction of the existence of a long-range  ${\langle p_T \rangle}$-$N_{\text{ch}}$ correlations.  The last ones might be observed in the event-by-event studies of observables measured in separated pseudorapidity intervals.

\subsubsection{Charged-particle multiplicity distribution}

\begin{figure}
  \includegraphics[height=.5\textheight]{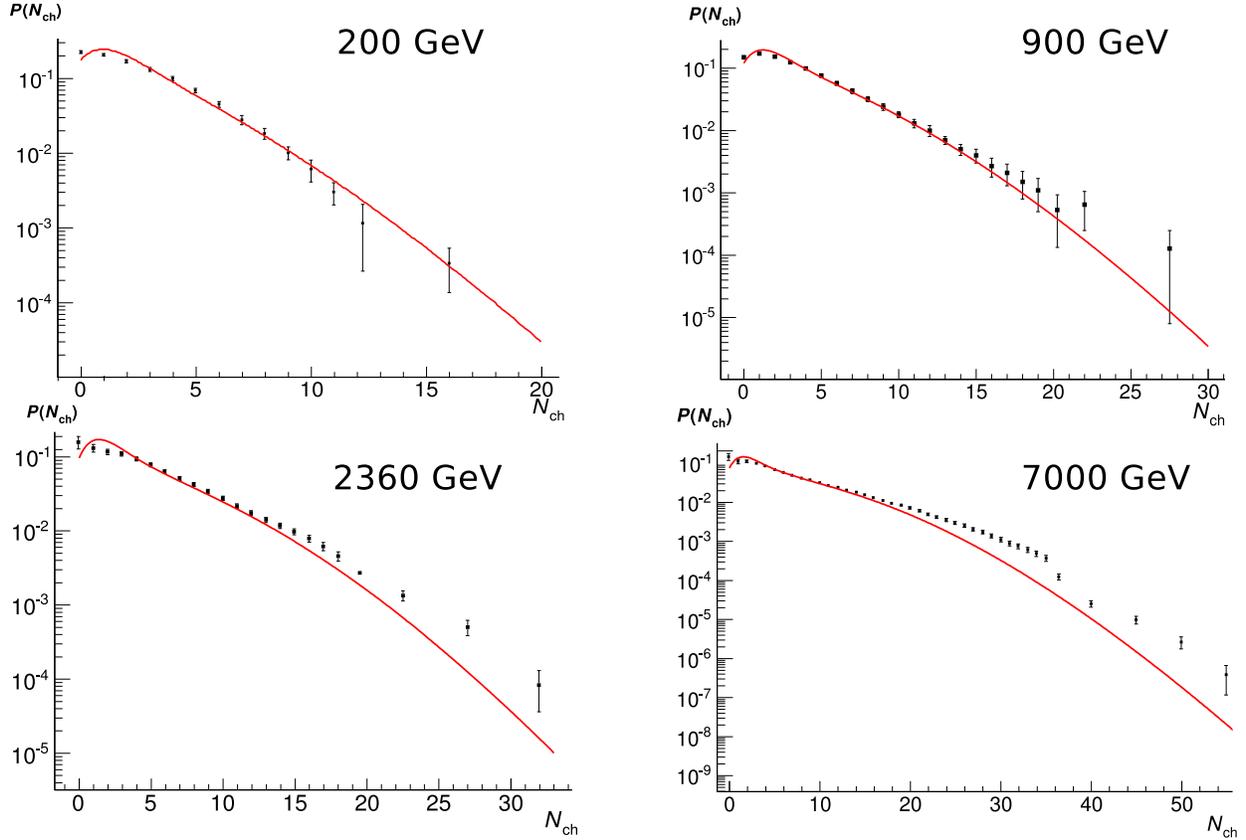}
  \label{mmd-4}
  \caption{ Charged particles multiplicity distributions  in $\pap$ collisions at various energies $\sqrt{s}$ ({200 GeV}, {900 GeV}) \cite{DistCMS} and in $pp$ collisions ({2360 GeV}, {7 TeV}) \cite{DistUA5}  in pseudorapidity interval $|\eta|<0.5$ (dots). The results of the  modified multi-pomeron exchange model  (lines) are calculated with the model parameters $\beta$ and $t$ as defined in  Fig. 1, 6.}
\end{figure}

%GF
The multiplicity distributions of charged particles formed as the result of processes going via the exchange of $n$  soft pomerons in $pp$ collisions at the given energy, calculated in the  modified multi-pomeron exchange model,  are compared to the relevant experimental data  in the Figure \ref{mmd-4}. Calculations were done using equation (\ref{e2}).
Examples are shown for the collision energies of  $\sqrt{s}$ =200, 900, 2360 and 7000 GeV.
% (The intervals of pseudorapidity are indicated in the figure caption -- NB!!!)   $\eta$ ....

We may conclude here that the overall tendencies of multiplicity distributions are well reproduced in wide energy range. The model predictions slightly
deviate from experimental data, that is observed in the region of the high-multiplicity tails,
%%GF
however this could be related both to some assumptions  used in the present calculations and to the still possible contributions of hard processes.

%from experimental data  observed in the region of the high-multiplicity tails are in general within about 1 error bar.

%The role of multi-pomeron processes in violation of the KNO-scale could be also shown here...(THIS IS DESIRABLE IN THE MAIN PAPER...)

\subsubsection{Energy dependence of ${\langle p_T \rangle}$ values}

\begin{figure}
  \includegraphics[height=.3\textheight]{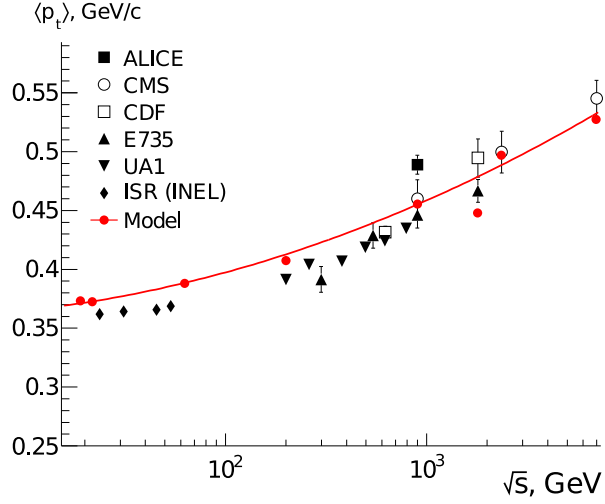}
   \label{mean-pt-energy}
   \caption{Energy dependence of mean $p_T$ values calculated in the modified multi-pomeron model with collectivity and comparison to the experiment. Data -- (black points) were borrowed from \cite{CMSptFit}.  Red points, %VKov
 approximated by a curve -- modified multi-pomeron exchange model with parameters $k$, $\beta$ and $t$ as in Fig. \ref{mult}, \ref{beta-t}.} %VKov
\end{figure}

Results of calculations in the modified multi-pomeron model with collectivity of the mean $p_T$ values  vs. $\sqrt{s}$  are compared to the experimental data  in the Figure  \ref{mean-pt-energy}. Calculations were done using equation (\ref{a9}).
The parameters of the model are the same as were defined above (see Fig. \ref{mult}, \ref{beta-t}).

\subsubsection{Energy dependence of string fusion model ratio $k/p_T^2$ }

\begin{figure}\label{kobt}
  \includegraphics[height=.3\textheight]{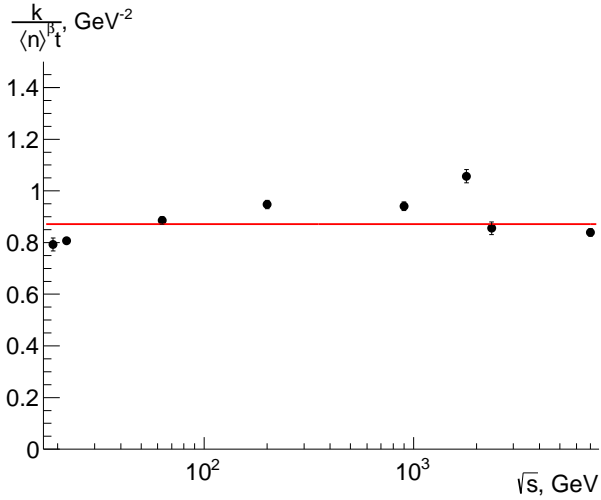}
  \caption{
%Picture 10
  Relation to string fusion hypothesis: ratio of $k$- mean multiplicity of charged particles from one string  to characteristic string tension at different $pp$ and $\pap$ collision energies. Dots -- experimental values from our analysis. Straight line -- approximation.}
\end{figure}

 We also made  checks of the string fusion model condition (\ref{b5}). Results are shown in the Fig. \ref{kobt} approximated by a straight line. Thus we can conclude here that the collectivity effects, relevant to quark-gluon string interaction phenomenon (string fusion), could be considered as important in $pp$ and $p\bar{p}$ collisions  and in a wide energy range, much wider than it was previously expected.

 \begin{equation}
   \label{b6}
   \frac{k}{{\langle n
   \rangle}^\beta t} = 0.87 \pm 0.08\,. %VKov
  \end{equation}

\section{Conclusions}
All features of ${\langle p_T \rangle}$-$N_{\text{ch}}$ correlations experimentally observed in the central rapidity region in $pp$ and $\pap$ collisions, in a wide energy range from the ISR to Tevatron and LHC, are successfully described in the framework of a new variant of the  multi-pomeron exchange model in which string collectivity has been included in an effective way.
The crucial feature of new model is the smooth logarithmic growth of mean multiplicity per string (the parameter $k$) with energy, whereas in previous works it was supposed to be constant. Thus the number of the model parameters is reduced  from three to two: $\beta-$parameter, responsible for the collective properties and $t-$ string tension. Both $\beta$ and $t$ are showing smooth energy dependence when  extracted from the available experimental data on ${\langle p_T \rangle}$-$N_{\text{ch}}$ correlations in  $pp$ and $\pap$ collisions.

The transition from negative to positive ${\langle p_T \rangle}$-$N_{\text{ch}}$
%VKov
 correlation pattern and tendency for flattening of the ${\langle p_T \rangle}$ with multiplicity and with increase of energy $\sqrt{s}$ from 17 GeV to 7000 GeV in $pp$ and $\pap$ collisions are quantitatively described within this single  approach. The experimental data on multiplicity distributions in the whole interval of collision energies and the  ${\langle p_T \rangle}$ vs. energy  dependence are also well reproduced numerically. Finally, the predictions of the string fusion model are also checked for the energy dependence of two model quantities - $k$ and $\beta$.
%%GF
All these results on ${\langle p_T \rangle}$-$N_{\text{ch}}$
%VKov
 correlation analysis in $pp$ and $\pap$  collisions at $\sqrt{s}$ from 17 GeV to 7 TeV are found to be in agreement with string fusion model hypothesis.

%%GF
%The multi-pomeron exchange model with collectivity effects  relevant to  string fusion phenomenon   is proved to be dominating in multiparticle  production and  important in $pp$ and $p\bar{p}$ collisions  and in a wide energy range.

%Predictions for LHC energy at 14 TeV are made.

%Predictions of $\langle p_T \rangle_{N_{\text{ch}}} - N_{\text{ch}}$
% correlations in separated  rapidity windows (long-range correlations) is also discussed.(???)
%
%
%
%Experimental results on mean pt and on pt-n correlation are summarized in a wide energy range.
%2-parameter Multi-pomeron exchange model is developed.
%
%of modified multi-pomeron exchange model with collectivity effects
%
%Smooth behavior of parameter $\beta$  	(responsible for string fusion)  with energy is observed.

%
%We propose a new variant of multi-pomeron model for the description of the correlations between
% mean transverse momentum and charged multiplicity in central rapidity window in $pp$ and $p\bar{p}$ collisions at $\sqrt{s}$ from 17 to 7000 GeV.
% We analyzed experimental data and obtained the dependence of the following model parameters on the energy: $k-$mean rapidity density
% of charged particles from one string, $\beta-$parameter, responsible for the collective properties and $t-$string tension.

%%%%%%%%%%%%%%%%%%%%%%%%%%%%%%%%%%%%%%%%%%%%%%%%
%% BACKMATTER
%%%%%%%%%%%%%%%%%%%%%%%%%%%%%%%%%%%%%%%%%%%%%%%%

\begin{theacknowledgments}
%%GF
 Authors are grateful to D.A.Derkach and V.V.Vechernin for useful discussions and permanent interest.  This work was partially supported by the SPbSU grant 11.38.66.2012. V.N.Kovalenko has also benefited from Special SPbSU Rector's Scholarship. %VKov
\end{theacknowledgments}



\bibliographystyle{aipproc}   % if natbib is available


\end{document}